\documentclass[aps,a4paper,twocolumn,groupedaddress]{revtex4-1}
\pdfoutput=1
\usepackage[]{latexsym,amsmath,amssymb,graphicx,epsfig}

\newcommand{\be}{\begin{eqnarray}}
\newcommand{\ee}{\end{eqnarray}}
\newcommand{\degree}{\ensuremath{^\circ}}

\begin{document}


\title{Back-to-Back Black Holes decay Signature at Neutrino Observatories}



\author{Nicusor Arsene}
\email[]{nicusorarsene@spacescience.ro}
\affiliation{Institute of Space Science, P.O.Box MG-23, Ro 077125 Bucharest-Magurele, Romania}
\affiliation{Physics Department, University of Bucharest, Bucharest-Magurele, Romania}

\author{Xavier Calmet}
\email[]{x.calmet@sussex.ac.uk}
\affiliation{Physics $\&$ Astronomy, University of Sussex,   Falmer, Brighton, BN1 9QH, UK }

\author{Laurentiu Ioan Caramete}
\email[]{lcaramete@spacescience.ro}
\affiliation{Institute of Space Science, P.O.Box MG-23, Ro 077125 Bucharest-Magurele, Romania }

\author{Octavian Micu}
\email[]{octavian.micu@spacescience.ro}
\affiliation{Institute of Space Science, P.O.Box MG-23, Ro 077125 Bucharest-Magurele, Romania }


\date{\today}

\begin{abstract}
We propose a decay signature for non-thermal small black holes with masses in the TeV range which can be discovered by neutrino observatories. The black holes would result due to the impact between ultra high energy neutrinos with nuclei in water or ice and decay instantaneously. They could be produced if the Planck scale is in the few TeV region and the highly energetic fluxes are large enough. Having masses close to the Planck scale, the typical decay mode for these black holes is into two particles emitted back-to-back. For a certain range of angles between the emitted particles and the center of mass direction of motion, it is possible for the detectors to measure separate muons having specific energies and their trajectories oriented at a large enough angle to prove that they are the result of a back-to-back decay event.
\end{abstract}


\maketitle

\section{Introduction}
In brane world models with a large extra-dimensional volume \cite{ArkaniHamed:1998rs, Antoniadis:1998ig, Randall:1999ee} or in even in four dimensions if there is  a large hidden sector of particles \cite{Calmet:2008tn}, quantum gravitational effects could become important anywhere between the traditional Planck scale, i.e. some $10^{16}$ TeV and a few TeV. If the Planck scale, i.e. the energy scale at which quantum gravitational effects become important, is in the lower end of this energy range, the collision of particles can result in the creation of small black holes with TeV masses when particles collide with  center of mass energies larger the Planck scale.

The formation of black holes in the collision of particles has been studied since  the 70's. In 1972, K.~Thorne proposed the {\em Hoop conjecture\/}~\cite{Thorne:1972ji} which states that  a black hole forms whenever the impact parameter $b$ of two colliding objects (of
negligible spatial extension) is shorter than the radius of the would-be-horizon
(roughly, the Schwarzschild radius, if angular momentum can be neglected)
corresponding to the total energy $M$ of the system \footnote{We
shall use units with $c=\hbar=1$ and the Boltzmann constant $k_{B}=1$,
and always display the Newton constant $G=l_{Pl}/M_{Pl}$, where $l_{Pl}$ and $M_{Pl}$
are the Planck length and mass, respectively.}
\begin{eqnarray}
b\lesssim \frac{2\,l_{Pl}\,M}{M_{Pl}}
\ .
\label{hoop}
\end{eqnarray}
While the hoop conjecture is intuitively very satisfactory, it is not enough to prove that black holes do indeed form in such collisions. However, there are now a proofs available of the formation of a closed trapped surface when two particles collider at very high energy above the Planck scale. The formation of a closed trapped surface suffices to demonstrate gravitational collapse and hence the formation of a black hole \footnote{The first calculations were performed by Penrose who never published his findings.}. The proofs \cite{D'Eath:1992hb, D'Eath:1992hd, D'Eath:1992qu, Eardley:2002re} covers both zero and non-zero impact parameters.  Remarkably the proof of Eardley and Giddings is analytical in the case of a four dimensional space-time \cite{Eardley:2002re}. This demonstrates the formation of a classical black hole in the collisions with a non-zero impact parameter of two particles with energies much larger than the Planck mass.  This work has been extended to the semi-classical regime by Hsu \cite{Hsu:2002bd}. Semi-classical black holes are expected to have  masses in the range from 5 to 20 times the Planck scale \cite{Meade:2007sz}.

Most of the articles related to the production of small black holes via particle collisions at colliders or in cosmic rays have considered semi-classical black holes 
\cite{Dimopoulos:2001hw, Banks:1999gd, Giddings:2001bu, Feng:2001ib, Anchordoqui:2003ug, Anchordoqui:2001cg, Anchordoqui:2003jr, Kowalski:2002gb, Ringwald:2001vk}. It is however possible that the center of mass energy available in such high energy collisions is not large enough to create semi-classical black holes. It was therefore proposed \cite{Calmet:2008dg, Calmet:2011ta, Calmet:2012cn} to consider quantum black hole which are non-thermal objects with masses close to the Planck mass which should be easier to produce. As they are non-thermal objects,  quantum black holes are expected to decay into a small number of particles, typically two. Experimental signatures for such decays are very different from the one of semi-classical objects which are expected to decay into several particles in a final explosion, see e.g. \cite{Cavaglia:2002si, Kanti:2004nr} for recent reviews. 

Bounds on the Planck scale using Earth skimming neutrinos creating black holes in the Earth crust have been derived in \cite{Anchordoqui:2003jr, Anchordoqui:2001cg, Calmet:2008rv}.
The back-to-back decay signature for quantum black holes produced in cosmic ray events was first proposed in \cite{Calmet:2012mf}. The authors study the possibility for the two particle showers produced by the particles resulting from the back-to-back decay of the black holes to be spatially separated and detected as two simultaneous shower events by cosmic ray observatories (current earth based or future space based experiments). That case study refers to quantum black holes which are generated due to the interaction of ultra high energy cosmic rays (UHECR) or neutrinos with particles in the upper atmosphere and decay instantaneously into two particles which move back-to-back in the center of mass reference frame. It is shown in \cite{Calmet:2012mf} that even if a small percentage of this type of events can be detected, there is parameter space for which detection is possible. 

Here we propose to take the idea one step further by analyzing whether it is possible to discover such a black hole decay signature in ice or water with the help of neutrino observatories. 
Besides being produced at colliders or in the atmosphere by high energetic collisions of cosmic rays with nuclei, quantum black holes can also be produced due to the collision between highly energetic neutrinos with nuclei in water or ice. Neutrino observatories are designed to detect muons induced by high-energy neutrinos. Because of the characteristics of the propagation of muons in water and ice, the neutrino direction can be derived with high accuracy \cite{2011NIMPA.656...11A,2006APh....26..155I}. As neutrinos only interact weakly and their trajectory points back to their sources, energetic neutrino events would point directly towards sources capable of producing these energetic events. We propose that these observatories can be used to search for quantum black hole events. Indeed, when neutrinos of high enough energies collide with particles in water or ice, if the center of mass energy is larger than the Planck mass, quantum black holes can be created. As stated before, black holes with masses close to the Planck mass decay preferentially into two particles which then produce secondary showers which can be seen by the neutrino experiments. 

If the Planck mass is of the order of a few TeV,  only black holes with masses above this energy scale can be produced. This implies that, to form a black hole, the energy of the neutrino has to be of the order of $10^7$GeV or above. The range of interaction lengths for neutrino energies ($E_{\nu} $) between $10^7$ - $10^9$ GeV is $6.6 \times 10^3$ - $9.4 \times 10^2$ km water equivalent in rock \cite{Gandhi:1995tf}. This means that the Earth is opaque to electron and muon neutrinos with energies in this range or larger. Only Earth skimming neutrinos \cite{Feng:2001ue} and those coming from above the horizon are thus useful for our considerations. 

\section {Black holes production} 
In this section we briefly describe the production cross section for quantum black hole formation. Note that the formulas are extrapolated from the semi-classical regime.  The black hole production cross section as a result of a neutrino interacting with nucleon ($\nu$ N $\to$ BH) is given by
\begin{eqnarray}
\label{CSection}
\sigma(E_\nu,x_{min},M_D)&=&
\int^1_0 2z dz \int^1_{\frac{( x_{min} M_D)^2}{y(z)^2 s_{max}}} dx F(n)  \\ && \nonumber \pi r_s^2(\sqrt{\hat s},M_{D}) \sum_i f_i(x,Q).
\end{eqnarray} 
 In this equation $M_D$ is the $4+n$ dimensional reduced Planck mass, $z=b/b_{max}$ with $b$ the impact parameter and $b_{max}$ the maximum value of the impact parameter for which black hole creation can occur as a result of the collision between the two particles, $x_{min}=M_{BH,min}/M_D$ and  $n$ is the number of extra-dimensions. $F(n)$ and $y(z)$ are the factors introduced by Eardley and Giddings \cite{Eardley:2002re} and by Yoshino and Nambu \cite{Yoshino:2002br}. The Schwarzschild radius in $4 + n$ dimensions is given by 
\begin{eqnarray}
r_s(us,n,M_D)=k(n)M_D^{-1}[\sqrt{us}/M_D]^{1/(1+n)}
\end{eqnarray}
where
\begin{eqnarray}
k(n) =  \left [2^n \sqrt{\pi}^{n-3} \frac{\Gamma((3+n)/2)}{2+n} \right ]^{1/(1+n)}.
\end{eqnarray}
Furthermore, note that $\hat{s}= 2 x m_N E_{\nu}$, with $m_N$ the  nuclei mass and $E_{\nu}$ the neutrino energy. The functions $f_i(x,Q)$ are the parton distribution functions.  Black hole production by cosmic neutrinos might be suppressed in comparison to the production rate from UHECRs \cite{Stojkovic:2005fx}. However this is a model dependent question. It is worth mentioning that although the parton level black hole cross section grows with energy to some power, the cross sections  at the nuclei level go quickly to zero because of the energy dependence of the parton distribution functions.

Only the flux of highly energetic neutrinos is relevant for the present case of study, flux which can be estimated by considering two sites of productions: at the source and between the source and the detection place, usually Earth. The production sites of the extragalactic UHECR, which include (AGN) and (GRBs), are also associated with the ones for neutrinos which are produced through pion decay in proton-proton or proton-photon interactions within the source \cite{Olinto:2000sa}  and the flux depends on the composition of the cosmic rays at high energies, which can be pure protons, neutrons, heavy nuclei or a mix of these \cite{Allard:2006mv,Hooper:2004jc}.

In order to estimate the number of  quantum black hole events expected at a neutrino detector like IceCube, we  use two models for the energy flux of the neutrino at high energy proposed in  \cite{2013arXiv13011703K}.
Here they use a a smoothly-broken power law to estimate the fluxes of high energy neutrinos reaching the Earth that also include the recent observation of two PeV-energy shower events by IceCube. These two models are based on a $E^{-2}$ accelerated proton spectrum, \cite{2001PhRvD63b3003M}, and use first a $\pi^+$ only decay channel and then  $\pi^{\pm}$ and $\mu^{\pm}$ decay channels to produce neutrinos. We combine this with the geometrical acceptance of the IceCUBE detector, \cite{2013NIMPA700188I} to find the numbers of the quantum black holes produced by a flux of neutrinos in the energy range between $10^7$ to $10^8$ GeV at IceCUBE in one year. This number also depends on  the model taken into consideration (i.e. the number of space-time dimensions, $(0,1,2,3,4,5,6,7)$ where $0$ corresponds to the large hidden sector model, $n=1$ to Randall Sundrum brane world model and higher $n$ models to ADD brane world models) and on the value of the Planck mass. Note that for estimating the number of black holes produced we will set $F(n)=1$ and $y(z)=1$. These factors are not known for non-thermal quantum black holes but they are expected to be order unity. Following \cite{Calmet:2012fv}, it is easy to estimate the branching ratios for the decomposition of non-thermal black holes. Assuming flavor conservation and baryon number conservation we find that black holes produced from a $u$-quark   and a muon-neutrino will decay 50 $\%$ of the time back to $\nu_\mu + u$ and  $\mu + d$. The same applies to black holes formed by the collision of a $d$-quark   and a muon-neutrino. As it will become obvious from the next two sections only a very small fraction of the back-to-back black hole decays will result in a clear signature. Tables \ref{table:one} and \ref{table:two} will give the number of events expected to be seen (after factoring in the branching ratio for the different decay channels and the percentage of the decays which will produce distinguishable signatures) by the IceCube neutrino observatory as functions of the Planck mass $M_{Pl}$ and the number of extra dimensions $n$.
 
\section {Back-to-back black holes decay signature in water and ice}

A black hole produced as a result of the collision between a highly energetic neutrino of energy $E_{\nu}$ (we shall neglect neutrino masses) and a particle of mass $m$ has a rest mass $M_{BH}$ equal to
\be
M_{BH}=\sqrt{m^2+2~E_{\nu}~ m}~,
\label{mbh}
\ee
and is moving relativistically with 
\be
\gamma_{BH}=\frac{E_{\nu}+m}{M_{BH}}~.
\label{pbh}
\ee
Obviously an on-shell black hole can form only when $M_{BH}>M_{Pl}$. In the remaining of this paper we shall assume that this condition is satisfied. Furthermore, we use $M_{BH}$ and $\gamma_{BH}$ to refer to the black hole mass and to the Lorentz factor. Note that the quantum black hole mass is determined by the neutrino energy and is a continuous quantity. For our calculation and our numerical simulations knowing the orders of magnitude for $M_{BH}$ and $\gamma_{BH}$ is sufficient. In principle one could take take into account the amount of energy which is radiated via gravitational radiation, the dependence of the horizon formation and black hole mass on the value of the impact parameter $b$ and so on. However, assuming that the impact parameter is small enough for a black hole to form, the mass and Lorentz factor of the black hole which forms can vary by less than an order of magnitude when considering all the above mentioned effects. 

Quantum black holes are non-thermal and they are expected to decay to a small number of particles, most likely two.
In the center of mass the two particles produced by the instantaneous decay of such a black hole are emitted back-to-back due to momentum conservation. There is no preferred direction in which the decay takes place, since the differential cross section is angular independent and the energy of the resulting decay products is restricted by energy conservation which translates into fact that the sum of the two masses  $m_a$ and $m_b$ has to be smaller than $M_{BH}$.
In the center of mass reference frame the momenta of the two particles are opposite vectors with magnitudes equal to
\be
\!\!\!p\!=\!\frac{\left[\left( M_{BH}^2-(m_a+m_b)^2\right)\!\!\left( M_{BH}^2-(m_a-m_b)^2\right)\right]^{\frac{1}{2}}}{2M_{BH}}.
\ee

The energies and momenta of the two particles in the laboratory reference frame (Earth reference frame) are
 \be
        \left(\begin{array}{c} E_i' \\ p_{i \parallel}' \end{array}\right) &=&
        \begin{pmatrix} \gamma_{BH} & -\beta_{BH} \gamma_{BH}  \\ -\beta_{BH} \gamma_{BH}  & \gamma_{BH} 
         \end{pmatrix}  \label{LT}
        \left(\begin{array}{c}  E_i \\ p_{i \parallel} \end{array}\right)\\
      p_{i \perp}' &=&p_{i \perp} \nonumber
    \ee
where $i=a, b$; $E_i$ and $p_i$ are the energy and momentum for the $i$-th particle measured in the center of mass, while the primed quantities are the corresponding ones measured in the reference frame of the Earth. In Eq. \ref{LT}, $p_{i\parallel}$ and $p_{i\perp}$ represent the momentum component parallel respectively perpendicular to the direction of motion of the center of mass.

One more ingredient is needed to carefully analyze the proposed black hole signature, which is the angle between the two showers in the Earth reference frame. One can start from $\phi_a$ and $\phi_b$, which are the angles between the two emitted particles in the center of mass reference frame measured from the direction of motion of the center of mass ($\phi_a + \phi_b=\pi$ since the particles are moving back-to-back) and perform a simple Lorentz transformation.  The resulting angles in the laboratory reference frame are
\be
\tan {\theta_i}=\frac{\sin{\phi_i}}{\gamma_{BH} \beta_{BH}\frac{E_i}{p_i}+\gamma_{BH} \cos{\phi_i}}~,
\label{tangent}
\ee 
These are the angles between the secondary showers and the direction of motion of the center of mass and the angle between the two showers is their sum.

\begin{figure}[t]
\includegraphics[scale=0.7]{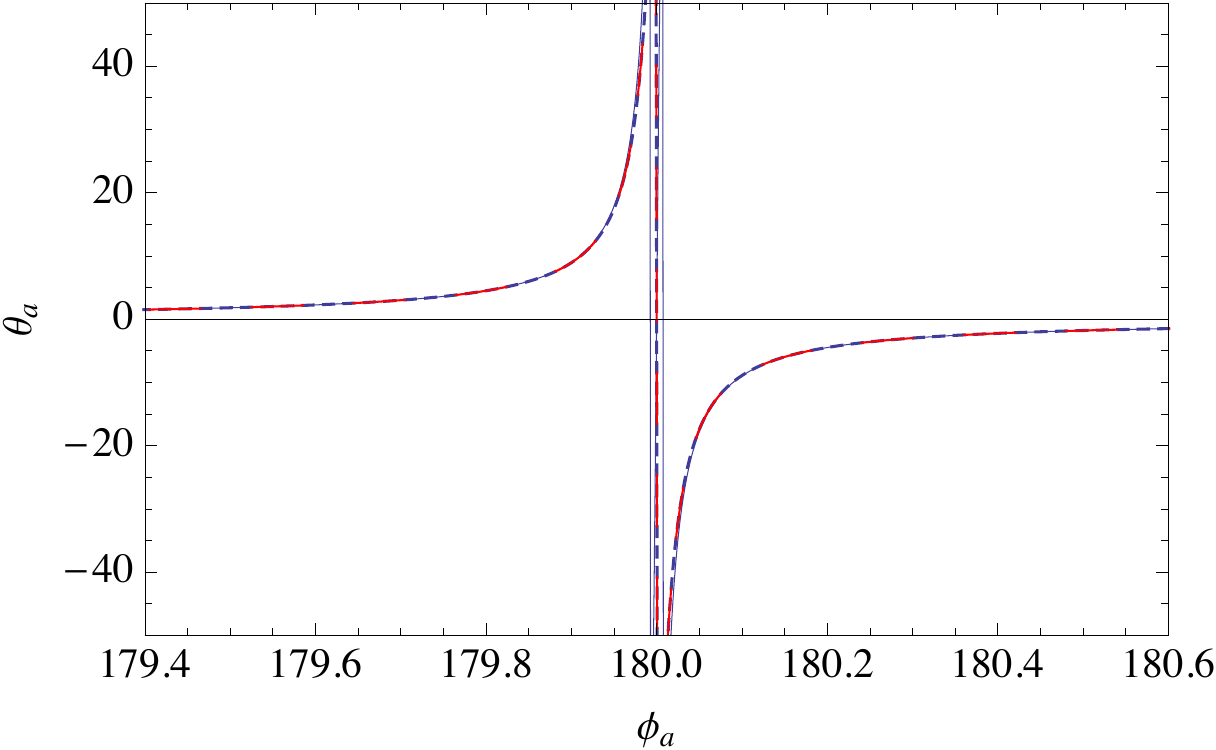}
\caption{Angle $\theta_a$ as  function of the angle $\phi_a$ for $m_a$ corresponding to: $m_{\pi^{+}}=139.5$ Mev and $m_{\pi^{-}}=139.5$ MeV blue solid line, $m_{p^+}=938.2$ MeV and $m_{\mu}= 105.7$ MeV long doted red line, $m_{p^+}=938.2$ MeV and $m_{\tau}= 1.777$ GeV small doted blue line, (decreasing from a larger possible angle for the lowest value of $m_a$, to a lower possible one for the largest value of $m_a$.} 
\label{plot1}
\end{figure} 

This distinctive black hole decay signature can be observed if the secondary showers are separated by an angle large enough for the experiment to be able to resolve the event into two distinctive coincident showers. This signature was already studied in detail for showers produced in the atmosphere in \cite{Calmet:2012mf}. As it is also specified in the reference cited above, numerical simulations show that there is parameter space for large angular separation between the two secondary showers. 
The plot in Fig. \ref{plot1} shows the angle of separation between the two secondary showers measured in the experiment reference frame ($\theta_a$) as a function of the corresponding angle in the center of mass reference frame ($\phi_a$). One notices that for a range of values of the angle $\phi_a$ covering about $0.4 \degree$ ($179.8\degree<\phi_a<180.2\degree$), $\theta_a$ takes large enough values for the showers to be resolved into two separate ones.

\begin{figure}[t]
\includegraphics[scale=0.18]{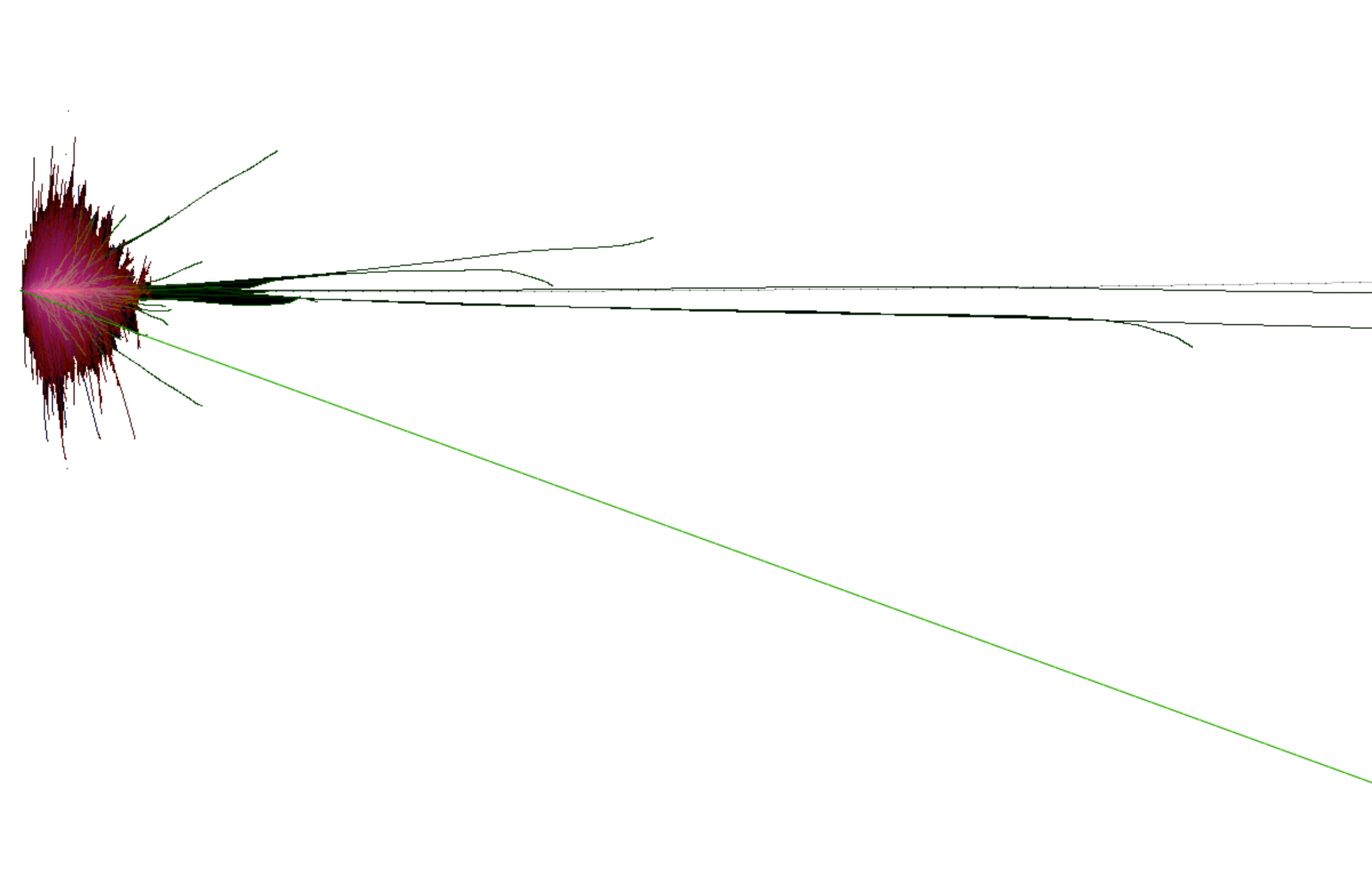}
\caption{CORSIKA simulation of a quantum black hole decay into a proton and a muon in water. The black hole is produced by a $10^{17}$ eV neutrino which collides with a neutron has a mass of approximately $10^{13}$ eV and decays immediately. In the Earth reference frame the muon has an energy of $3\times 10^{11}$ eV and the proton has an energy of approximately $10^{17}$ eV. The angle between the two particles in the Earth reference frame is $5\degree$.} 
\label{plot2}
\end{figure}

The image in Fig. \ref{plot2} shows the simulation of an approximately 10 TeV quantum black hole decay into a muon and a proton ($BH\to \mu^- +p^+$) in water. Such a black hole can be produced via a collision between a $10^{17}$ eV neutrino with a nucleon at rest. Assuming a Planck mass in the few TeV range the black hole mass is just above the Planck scale and it decays preferentially into two particles (for the Planck mass anywhere above 2 TeV the black hole mass is less than five times the value of the Planck mass). Strictly speaking the black hole is formed by the scattering between the neutrino and a parton from the the nuclei. However for the sake of the simulation is it best to consider the nuclei instead of its constituents. Needless to say, for all charges to be conserved it was assumed that the black hole formed as a result of the impact between a muon neutrino and a neutron ($\nu_{\mu}+n\to BH$). It needs to be pointed out that this is just one of the possible decay channels for quantum black holes and it was chosen because it is suitable for the detection of the back-to-back decay signature (we do not mean to imply that this is the only decay channel for which the signature can be discovered). The numerical simulations were performed using CORSIKA-6600-WI-0.9 \footnote{This version of CORSIKA was obtained changing the medium in which the simulation takes place from air to ice. It is maintained by J. Bolmont in DESY at Zeuthen. This work was inspired by the work done by T. Sloan (Lancaster University) for the ACoRNE collaboration (Astropart. Phys., 28, 366 (2007)).}, (COsmic Ray SImulations for KAscade, a software package developed to perform detailed simulation of extensive air showers initiated by high energy cosmic ray particles - a version of which was also extended to simulate particle showers which develop in ice or water) \cite{corsika, corsika1}. More specifically the QGSJET 01C model for hadronic interactions at ultra-high energies was used \cite{1997NuPhS..52...17K}. 

In the Earth reference frame the energies of the two black hole decay products are approximately $10^{17}$ eV for the proton and $3\times 10^{11}$ eV for the muon. The reason for the large difference between the two energies comes from the fact that the angle in the laboratory reference frame can only take large values when the angles in the center of mass frame are close to $0\degree$ respectively $180\degree$ (see Fig. \ref{plot1}). The resulting large but opposite values of $p_{\parallel}$ for the two particles, combined with the large Lorentz factor of the center of mass, leads to the several orders of magnitude difference between the two energies. An angle of $5\degree$ between the two particles in the Earth reference frame was used for this simulation (experiments with large enough resolution capabilities can possibly detect even smaller angles). For the sake of illustration, the axes of the plot have different scales, the angle thus appears larger that it really is. For the angle considered here, after the muons travel a distance of about a kilometer the separation between the single muon and the bunch of muons remaining from the proton shower is about 100 meters. 
 One can see that the muon track (represented in green) is very well separated from the bunch of muons (black muon tracks) resulting from the shower produced by the proton, therefore the muons can be identified as coming from separate showers but having the same origin. When analyzing the numerical simulations one also notices that most of the proton shower dies out very fast with the only particles propagating beyond a few tens of meters being a bunch of energetic muons. The muons can be identified by the Cherenkov light which is emitted by the particles traveling through water or ice. The situation is very similar for events taking place in ice and this is obvious if when comparing the plots in Figs. \ref{water1} and \ref{ice1}.

\begin{widetext}

\begin{figure}[t]
\centerline{\begin{tabular}{ r}
\centerline{\hbox{
\epsfig{figure=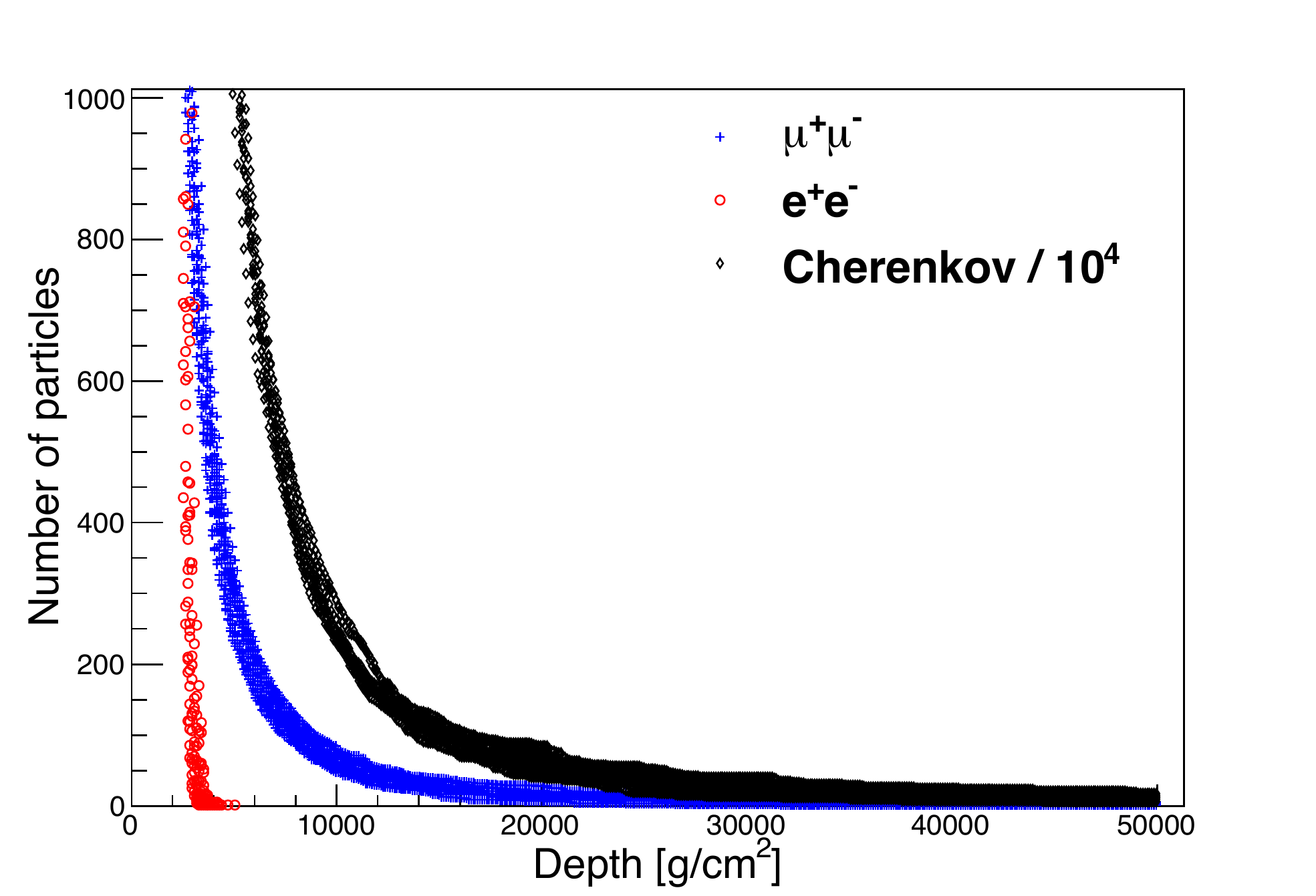,height=6cm,clip=}
\epsfig{figure=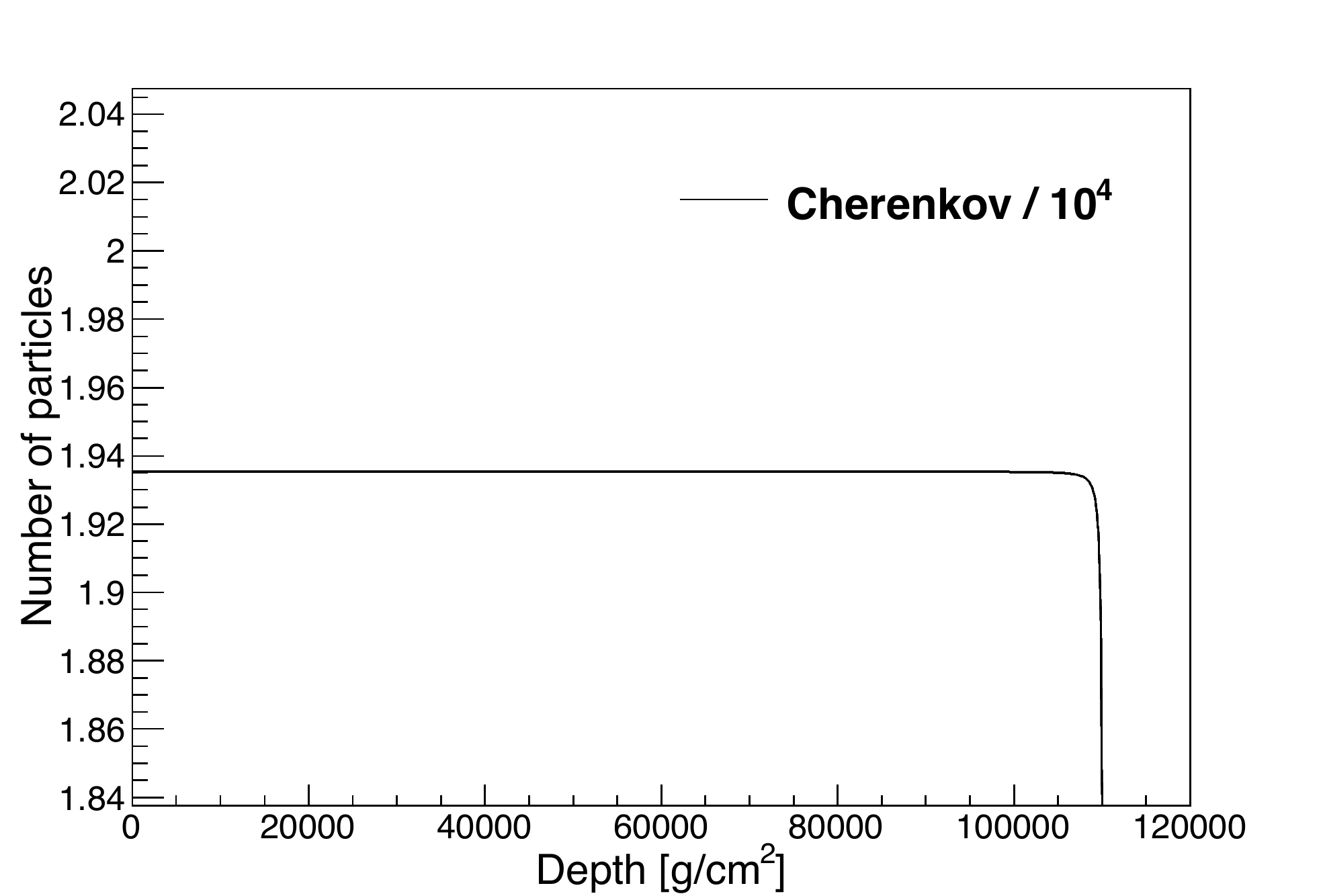,height=6cm,clip=}
}}
\end{tabular}}
\caption{Plots representing the number of particles in the showers as a function of the distance traveled through water. Each plot contains ten overlapping numerical simulations. The left panel shows the particles and Cherenkov light produced in a $10^{17}$ eV proton shower, while the right panel shows the Cherenkov light produced by the $3\times 10^{11}$ eV muon.}
\label{water1}
\end{figure}



\begin{figure}
\centerline{\begin{tabular}{ r}
\centerline{\hbox{
\epsfig{figure=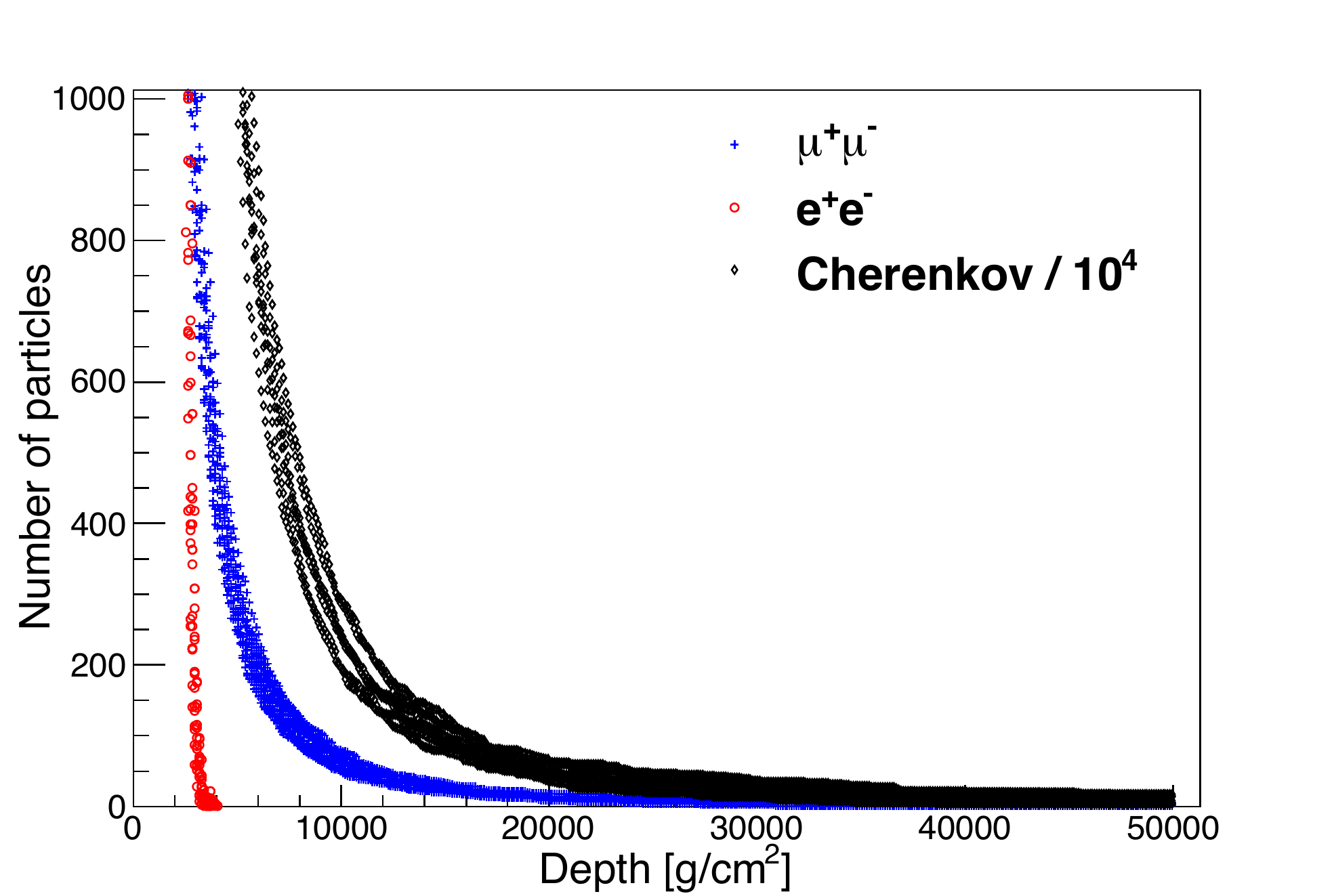,height=6cm,clip=}
\epsfig{figure=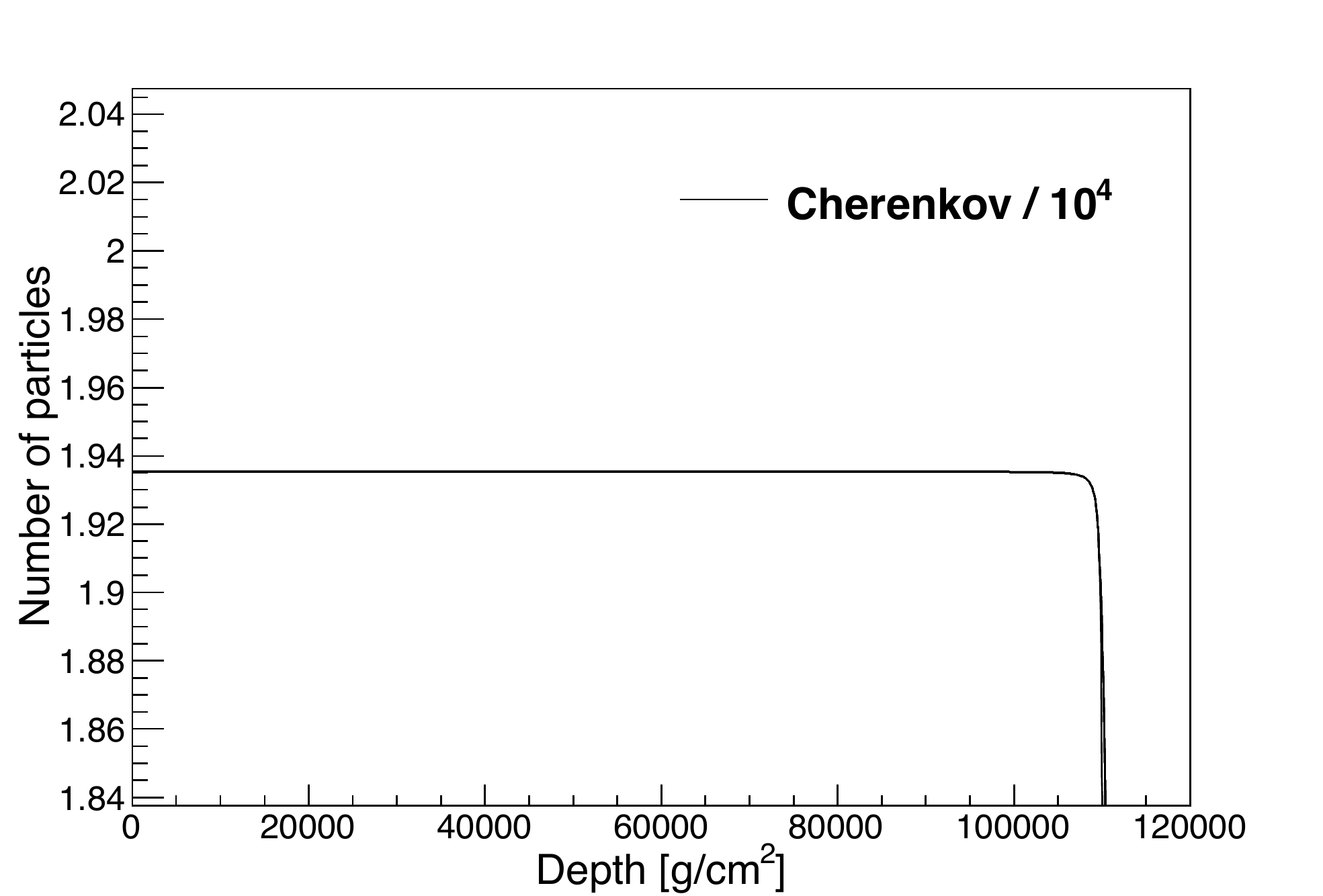,height=6cm,clip=}
}}
\end{tabular}}
\caption{Plots representing the number of particles in the showers as a function of the distance traveled through ice. Each plot contains ten overlapping numerical simulations. The left panel shows the particles and Cherenkov light produced in a $10^{17}$ eV proton shower, while the right panel shows the Cherenkov light produced by the $3\times 10^{11}$ eV muon.}
\label{ice1}
\end{figure}

\end{widetext}

Figs. \ref{water1} and \ref{ice1} show side by side the number of particles in the showers produced in water (respectively ice) by $10^{17}$ eV protons (plots on the left side) and $3\times 10^{11}$ eV muons (plots on the right side). 
The figures on the left show that the only remaining components of the proton showers beyond some $30$ meters are muons and Cherenkov light; while the hadronic and electromagnetic parts die out. The muons travel to distances of the order of kilometers while continuously generating Cherenkov radiation which can be measured by the underwater detectors, respectively by the detectors placed in ice. The figures on the right show the Cherenkov light produced by the single muons, which are in each case one of the two products resulting from the back-to-back decay of the quantum black holes. A $3\times 10^{11}$ eV muon travels a distance of the order of one kilometer in water (similar distances are traveled by muons moving through ice) and its track can be detected by the Cherenkov light which it produces. The muons remaining from the $10^{17}$ eV proton travel distances up to about two kilometers both in ice or water (the plots in Fig. \ref{water1} and \ref{ice1} only show the shower development for the first five hundred meters in order for the electromagnetic component to be visible). Beyond around on hundred meters, the number of Cherenkov photons emitted by the $3\times 10^{11}$ eV muons is of the same order of magnitude as the number of Cherenkov photons emitted by the muons generated in the $10^{17}$ eV proton showers. 

As mentioned earlier,  the original muon and the proton which generates the several other muons have very particular energies for a certain initial neutrino energy. This is a direct consequence of the small range of angles in the center of mass reference frame for which the angle in the laboratory reference frame can have large enough values for the muon tracks to be spatially separated. The IceCube collaboration specifies that they can separate track-like events with a very high accuracy. They can distinguish muon tracks separated by less than $1\degree$ (the collaboration claims $0.7\degree$).\cite{Karle:2010xx} When analyzing an event of the type described above one can also use the reconstructed energies to pinpoint back-to-back quantum black hole decay events.  The price to pay is that only about $5\times 10^{-3}~ \%$ of the total number of quantum black hole decay events can be discovered in this way. 

Measuring muon tracks with the characteristics described above in the data collected by a neutrino observatory such as IceCube or the future KM3NeT \cite{KM3NeT} would signal a back-to-back black hole decay event. 

Factoring in the percentage of the total number of events which can be observed ( $5\times 10^{-3}~ \%$) and the branching ratios one calculates the number of events expected to be seen per year both for the first model of the neutrino flux (Table \ref{table:one}) and for the second model (Table \ref{table:two}). The two tables contain the number of events which are expected to be seen by the IceCube experiment for the Planck mass varying from 1 to 8 TeV and the number extra dimensions $n=0,1,2,3,4,5,6,7$. As it can be seen from the tables, the number of expected events varies with the Plank mass and the actual scenario considered, but in many cases the numbers are large enough for the collaboration to see such events and eventually discover TeV scale gravity.

In the near future this numbers will increase by an order of magnitude, as the KM3NeT neutrino telescope (in construction) will be significantly larger.  It will have a volume of 4km$^{3}$ in each of its 3 locations (compared to IceCube which is a km-scale neutrino detector \cite{Aartsen:2013dla}), it will exceed IceCube in sensitivity and will also complement its field of view \cite{2013NIMPA.725...13K}.

\begin{widetext}

\begin{table}
\begin{tabular}{||c|c|c|c|c|c|c|c|c||}
\hline\hline
No. of extra & $M_{Pl}$ & $M_{Pl}$ & $M_{Pl}$ & $M_{Pl}$ & $M_{Pl}$ & $M_{Pl}$ & $M_{Pl}$ & $M_{Pl}$\\
dimensions & 1  TeV & 2 TeV & 3 TeV & 4 TeV  & 5 TeV  & 6 TeV  & 7 TeV  & 8 TeV\\ \hline\hline
0&0.11&$0.55\times 10^{-2}$&$0.80\times 10^{-3}$&$0.20\times 10^{-3}$&$0.10\times 10^{-3}$&$0.16\times 10^{-4}$&$0.52\times 10^{-5}$&$0.16\times 10^{-5}$\\ \hline
1&2.9&0.19&$0.33\times 10^{-1}$&$0.85\times 10^{-2}$&$0.26\times 10^{-2}$&$0.88\times 10^{-3}$&$0.30\times 10^{-3}$&$0.96\times 10^{-4}$\\ \hline
2&11&0.79&0.14&$0.40\times 10^{-1}$&$0.12\times 10^{-1}$&$0.41\times 10^{-2}$&$0.14\times 10^{-2}$&$0.46\times 10^{-3}$\\ \hline
3&25&1.8&0.33&$0.90\times 10^{-1}$&$0.29\times 10^{-1}$&$0.10\times 10^{-1}$&$0.35\times 10^{-2}$&$0.11\times 10^{-2}$\\ \hline
4&43&3.2&0.59&0.16&$0.51\times 10^{-1}$&$0.18\times 10^{-1}$&$0.62\times 10^{-2}$&$0.20\times 10^{-2}$\\ \hline
5&64&4.8&0.89&0.24&$0.78\times 10^{-1}$&$0.27\times 10^{-1}$&$0.95\times 10^{-2}$&$0.31\times 10^{-2}$\\ \hline
6&88&6.5&1.2&0.33&0.11&$0.38\times 10^{-1}$&$0.13\times 10^{-1}$&$0.44\times 10^{-2}$\\ \hline
7&110&8.5&1.6&0.44&0.14&$0.50\times 10^{-1}$&$0.17\times 10^{-1}$&$0.57\times 10^{-2}$\\ \hline
\hline
\end{tabular}
\caption{Number of black hole events per year expected at the IceCube experiment for which the separation of the two showers is larger than $1\degree$ in the reference frame of the laboratory/experiment when using the first model for the neutrino flux.}
\label{table:one}
\end{table}

\begin{table}
\begin{tabular}{||c|c|c|c|c|c|c|c|c||}
\hline\hline
No. of extra & $M_{Pl}$ & $M_{Pl}$ & $M_{Pl}$ & $M_{Pl}$ & $M_{Pl}$ & $M_{Pl}$ & $M_{Pl}$ & $M_{Pl}$\\
dimensions & 1  TeV & 2 TeV & 3 TeV & 4 TeV  & 5 TeV  & 6 TeV  & 7 TeV  & 8 TeV\\ \hline\hline
0&$0.30\times 10^{-1}$&$0.14\times 10^{-2}$&$0.20\times 10^{-3}$&$0.40\times 10^{-4}$&$0.10\times 10^{-4}$&$0.41\times 10^{-5}$&$0.13\times 10^{-5}$&$0.41\times 10^{-6}$\\ \hline
1&0.41&$0.27\times 10^{-1}$&$0.47\times 10^{-2}$&$0.12\times 10^{-2}$&$0.40\times 10^{-3}$&$0.12\times 10^{-3}$&$0.42\times 10^{-4}$&$0.13\times 10^{-4}$\\ \hline
2&1.4&0.10&$0.18\times 10^{-1}$&$0.47\times 10^{-2}$&$0.15\times 10^{-2}$&$0.51\times 10^{-3}$&$0.17\times 10^{-3}$&$0.57\times 10^{-4}$\\ \hline
3&2.9&0.21&$0.40\times 10^{-1}$&$0.10\times 10^{-1}$&$0.34\times 10^{-2}$&$0.12\times 10^{-2}$&$0.40\times 10^{-3}$&$0.13\times 10^{-3}$\\ \hline
4&4.9&0.35&$0.66\times 10^{-1}$&$0.18\times 10^{-1}$&$0.58\times 10^{-2}$&$0.20\times 10^{-2}$&$0.70\times 10^{-3}$&$0.23\times 10^{-3}$\\ \hline
5&7.1&0.52&0.10&$0.26\times 10^{-1}$&$0.86\times 10^{-2}$&$0.30\times 10^{-2}$&$0.10\times 10^{-2}$&$0.34\times 10^{-3}$\\ \hline
6&9.5&0.70&0.13&$0.36\times 10^{-1}$&$0.12\times 10^{-1}$&$0.41\times 10^{-2}$&$0.14\times 10^{-2}$&$0.47\times 10^{-3}$\\ \hline
7&12&0.90&0.17&$0.47\times 10^{-1}$&$0.15\times 10^{-1}$&$0.53\times 10^{-2}$&$0.18\times 10^{-2}$&$0.61\times 10^{-3}$\\ \hline
\hline
\end{tabular}
\caption{Number of black hole events per year expected at the IceCube experiment for which the separation of the two showers is larger than $1\degree$ in the reference frame of the laboratory/experiment when using the second model for the neutrino flux.}
\label{table:two}
\end{table}

\end{widetext}
\section {Conclusions}

We have described a novel approach to probe the possibility for the Plank mass if it is in the few TeV range via back-to-back decays of Planck scale quantum black holes using the data collected by neutrino observatories. 

This signature consists in detecting the Cherenkov radiation of two simultaneous muon tracks (to be more precise one of the tracks is made up of several muons while the other would be from a single original muon) oriented at an angle, but pointing to a common origin. Another particularity which makes this signature unique is the energy in the laboratory reference frame of the two products of the back-to-back decay of the black hole. In the case studied here a $10^{17}$ eV muon neutrino produces a $10^{13}$ eV black hole by colliding with a neutron. The quantum black hole can be discovered only when the two particles resulting from its back-to-back decay are almost aligned (in the center of mass reference frame) with the direction of motion of the center of mass. In this case the energies of the two decay products (muon and proton) in the Earth reference frame are approximately $10^{17}$ eV and $3\times 10^{11}$ eV; and the resulting muons produce two distinguishable Cherenkov light tracks at an angle of one to several degrees. 

The available parameter space for this signature to be discovered is very small (this, of course, depends on the high energy neutrino flux), but it is a very unique signature and therefore worth being considered. 


This black hole decay signature allows the neutrino observatories to perform entirely independent searches for the Planck scale, joining this way cosmic rays experiments and the LHC in the efforts to search for TeV scale micro black holes. 

\vspace{1mm}
\paragraph*{Acknowledgements:} This work is supported in part by the European Cooperation in Science and Technology (COST) action MP0905 ``Black Holes in a Violent  Universe". N.A., L.I.C. and O.M. were supported by research grants: UEFISCDI project PN-II-RU-TE-2011-3-0184 and LAPLAS 3. The work of X.C. is supported in part by the Science and Technology Facilities Council (grant number ST/J000477/1)

\bibliography{theorybib}

\providecommand{\noopsort}[1]{}\providecommand{\singleletter}[1]{#1}%
\begin{thebibliography}{10}%
\makeatletter
\providecommand \@ifxundefined [1]{%
 \ifx #1\undefined \expandafter \@firstoftwo
 \else \expandafter \@secondoftwo
\fi
}%
\providecommand \@ifnum [1]{%
 \ifnum #1\expandafter \@firstoftwo
 \else \expandafter \@secondoftwo
\fi
}%
\providecommand \enquote [1]{``#1''}%
\providecommand \bibnamefont  [1]{#1}%
\providecommand \bibfnamefont [1]{#1}%
\providecommand \citenamefont [1]{#1}%
\providecommand\href[0]{\@sanitize\@href}%
\providecommand\@href[1]{\endgroup\@@startlink{#1}\endgroup\@@href}%
\providecommand\@@href[1]{#1\@@endlink}%
\providecommand \@sanitize [0]{\begingroup\catcode`\&12\catcode`\#12\relax}%
\@ifxundefined \pdfoutput {\@firstoftwo}{%
 \@ifnum{\z@=\pdfoutput}{\@firstoftwo}{\@secondoftwo}%
}{%
 \providecommand\@@startlink[1]{\leavevmode\special{html:<a href="#1">}}%
 \providecommand\@@endlink[0]{\special{html:</a>}}%
}{%
 \providecommand\@@startlink[1]{%
  \leavevmode
  \pdfstartlink
   attr{/Border[0 0 1 ]/H/I/C[0 1 1]}%
   user{/Subtype/Link/A<</Type/Action/S/URI/URI(#1)>>}%
  \relax
 }%
 \providecommand\@@endlink[0]{\pdfendlink}%
}%
\providecommand \url  [0]{\begingroup\@sanitize \@url }%
\providecommand \@url [1]{\endgroup\@href {#1}{\urlprefix}}%
\providecommand \urlprefix [0]{URL }%
\providecommand \Eprint[0]{\href }%
\@ifxundefined \urlstyle {%
  \providecommand \doi [1]{doi:\discretionary{}{}{}#1}%
}{%
  \providecommand \doi [0]{doi:\discretionary{}{}{}\begingroup
  \urlstyle{rm}\Url }%
}%
\providecommand \doibase [0]{http://dx.doi.org/}%
\providecommand \Doi[1]{\href{\doibase#1}}%
\providecommand \bibAnnote [3]{%
  \BibitemShut{#1}%
  \begin{quotation}\noindent
    \textsc{Key:}\ #2\\\textsc{Annotation:}\ #3%
  \end{quotation}%
}%
\providecommand \bibAnnoteFile [2]{%
  \IfFileExists{#2}{\bibAnnote {#1} {#2} {\input{#2}}}{}%
}%
\providecommand \typeout [0]{\immediate \write \m@ne }%
\providecommand \selectlanguage [0]{\@gobble}%
\providecommand \bibinfo [0]{\@secondoftwo}%
\providecommand \bibfield [0]{\@secondoftwo}%
\providecommand \translation [1]{[#1]}%
\providecommand \BibitemOpen[0]{}%
\providecommand \bibitemStop [0]{}%
\providecommand \bibitemNoStop [0]{.\EOS\space}%
\providecommand \EOS [0]{\spacefactor3000\relax}%
\providecommand \BibitemShut [1]{\csname bibitem#1\endcsname}%
\bibitem{ArkaniHamed:1998rs}%
  \BibitemOpen
  \bibfield{author}{%
  \bibinfo {author} {\bibfnamefont{N.}~\bibnamefont{Arkani-Hamed}}, \bibinfo
  {author} {\bibfnamefont{S.}~\bibnamefont{Dimopoulos}},\ and\ \bibinfo
  {author} {\bibfnamefont{G.}~\bibnamefont{Dvali}},\ }%
  \bibfield{journal}{%
  \Doi{10.1016/S0370-2693(98)00466-3}{\bibinfo {journal} {Phys.Lett.}}\ }%
  \textbf{\bibinfo {volume} {B429}},\ \bibinfo {pages} {263} (\bibinfo {year}
  {1998}),\ \Eprint{http://arxiv.org/abs/hep-ph/9803315}{arXiv:hep-ph/9803315
  [hep-ph]}%
  \bibAnnoteFile{NoStop}{ArkaniHamed:1998rs}%
\bibitem{Antoniadis:1998ig}%
  \BibitemOpen
  \bibfield{author}{%
  \bibinfo {author} {\bibfnamefont{I.}~\bibnamefont{Antoniadis}}, \bibinfo
  {author} {\bibfnamefont{N.}~\bibnamefont{Arkani-Hamed}}, \bibinfo {author}
  {\bibfnamefont{S.}~\bibnamefont{Dimopoulos}},\ and\ \bibinfo {author}
  {\bibfnamefont{G.}~\bibnamefont{Dvali}},\ }%
  \bibfield{journal}{%
  \Doi{10.1016/S0370-2693(98)00860-0}{\bibinfo {journal} {Phys.Lett.}}\ }%
  \textbf{\bibinfo {volume} {B436}},\ \bibinfo {pages} {257} (\bibinfo {year}
  {1998}),\ \Eprint{http://arxiv.org/abs/hep-ph/9804398}{arXiv:hep-ph/9804398
  [hep-ph]}%
  \bibAnnoteFile{NoStop}{Antoniadis:1998ig}%
\bibitem{Randall:1999ee}%
  \BibitemOpen
  \bibfield{author}{%
  \bibinfo {author} {\bibfnamefont{L.}~\bibnamefont{Randall}}\ and\ \bibinfo
  {author} {\bibfnamefont{R.}~\bibnamefont{Sundrum}},\ }%
  \bibfield{journal}{%
  \Doi{10.1103/PhysRevLett.83.3370}{\bibinfo {journal} {Phys.Rev.Lett.}}\ }%
  \textbf{\bibinfo {volume} {83}},\ \bibinfo {pages} {3370} (\bibinfo {year}
  {1999}),\ \Eprint{http://arxiv.org/abs/hep-ph/9905221}{arXiv:hep-ph/9905221
  [hep-ph]}%
  \bibAnnoteFile{NoStop}{Randall:1999ee}%
\bibitem{Calmet:2008tn}%
  \BibitemOpen
  \bibfield{author}{%
  \bibinfo {author} {\bibfnamefont{X.}~\bibnamefont{Calmet}}, \bibinfo {author}
  {\bibfnamefont{S.~D.}\ \bibnamefont{Hsu}},\ and\ \bibinfo {author}
  {\bibfnamefont{D.}~\bibnamefont{Reeb}},\ }%
  \bibfield{journal}{%
  \Doi{10.1103/PhysRevD.77.125015}{\bibinfo {journal} {Phys.Rev.}}\ }%
  \textbf{\bibinfo {volume} {D77}},\ \bibinfo {pages} {125015} (\bibinfo {year}
  {2008}),\ \Eprint{http://arxiv.org/abs/0803.1836}{arXiv:0803.1836 [hep-th]}%
  \bibAnnoteFile{NoStop}{Calmet:2008tn}%
\bibitem{Thorne:1972ji}%
  \BibitemOpen
  \bibfield{author}{%
  \bibinfo {author} {\bibfnamefont{K.~S.}\ \bibnamefont{{Thorne}}},\ }%
  \enquote{\bibinfo {title} {{Nonspherical Gravitational Collapse--A Short
  Review}},}\ in\ \emph{\bibinfo {booktitle} {Magic Without Magic: John
  Archibald Wheeler}},\ \bibinfo {editor} {edited by\ \bibinfo {editor}
  {\bibfnamefont{J.~R.}\ \bibnamefont{{Klauder}}}}\ (\bibinfo {year} {1972})\
  p.\ \bibinfo {pages} {231}%
  \bibAnnoteFile{NoStop}{Thorne:1972ji}%
\bibitem{Note1}%
  \BibitemOpen
  \bibinfo {note} {We shall use units with $c=\hbar =1$ and the Boltzmann
  constant $k_{B}=1$, and always display the Newton constant $G=l_{Pl}/M_{Pl}$,
  where $l_{Pl}$ and $M_{Pl}$ are the Planck length and mass, respectively.}%
  \bibAnnoteFile{Stop}{Note1}%
\bibitem{Note2}%
  \BibitemOpen
  \bibinfo {note} {The first calculations were performed by Penrose who never
  published his findings.}%
  \bibAnnoteFile{Stop}{Note2}%
\bibitem{D'Eath:1992hb}%
  \BibitemOpen
  \bibfield{author}{%
  \bibinfo {author} {\bibfnamefont{P.}~\bibnamefont{D'Eath}}\ and\ \bibinfo
  {author} {\bibfnamefont{P.}~\bibnamefont{Payne}},\ }%
  \bibfield{journal}{%
  \Doi{10.1103/PhysRevD.46.658}{\bibinfo {journal} {Phys.Rev.}}\ }%
  \textbf{\bibinfo {volume} {D46}},\ \bibinfo {pages} {658} (\bibinfo {year}
  {1992})%
  \bibAnnoteFile{NoStop}{D'Eath:1992hb}%
\bibitem{D'Eath:1992hd}%
  \BibitemOpen
  \bibfield{author}{%
  \bibinfo {author} {\bibfnamefont{P.}~\bibnamefont{D'Eath}}\ and\ \bibinfo
  {author} {\bibfnamefont{P.}~\bibnamefont{Payne}},\ }%
  \bibfield{journal}{%
  \Doi{10.1103/PhysRevD.46.675}{\bibinfo {journal} {Phys.Rev.}}\ }%
  \textbf{\bibinfo {volume} {D46}},\ \bibinfo {pages} {675} (\bibinfo {year}
  {1992})%
  \bibAnnoteFile{NoStop}{D'Eath:1992hd}%
\bibitem{D'Eath:1992qu}%
  \BibitemOpen
  \bibfield{author}{%
  \bibinfo {author} {\bibfnamefont{P.}~\bibnamefont{D'Eath}}\ and\ \bibinfo
  {author} {\bibfnamefont{P.}~\bibnamefont{Payne}},\ }%
  \bibfield{journal}{%
  \Doi{10.1103/PhysRevD.46.694}{\bibinfo {journal} {Phys.Rev.}}\ }%
  \textbf{\bibinfo {volume} {D46}},\ \bibinfo {pages} {694} (\bibinfo {year}
  {1992})%
  \bibAnnoteFile{NoStop}{D'Eath:1992qu}%
\bibitem{Eardley:2002re}%
  \BibitemOpen
  \bibfield{author}{%
  \bibinfo {author} {\bibfnamefont{D.~M.}\ \bibnamefont{Eardley}}\ and\
  \bibinfo {author} {\bibfnamefont{S.~B.}\ \bibnamefont{Giddings}},\ }%
  \bibfield{journal}{%
  \Doi{10.1103/PhysRevD.66.044011}{\bibinfo {journal} {Phys.Rev.}}\ }%
  \textbf{\bibinfo {volume} {D66}},\ \bibinfo {pages} {044011} (\bibinfo {year}
  {2002}),\ \Eprint{http://arxiv.org/abs/gr-qc/0201034}{arXiv:gr-qc/0201034
  [gr-qc]}%
  \bibAnnoteFile{NoStop}{Eardley:2002re}%
\bibitem{Hsu:2002bd}%
  \BibitemOpen
  \bibfield{author}{%
  \bibinfo {author} {\bibfnamefont{S.~D.}\ \bibnamefont{Hsu}},\ }%
  \bibfield{journal}{%
  \Doi{10.1016/S0370-2693(03)00012-1}{\bibinfo {journal} {Phys.Lett.}}\ }%
  \textbf{\bibinfo {volume} {B555}},\ \bibinfo {pages} {92} (\bibinfo {year}
  {2003}),\ \Eprint{http://arxiv.org/abs/hep-ph/0203154}{arXiv:hep-ph/0203154
  [hep-ph]}%
  \bibAnnoteFile{NoStop}{Hsu:2002bd}%
\bibitem{Meade:2007sz}%
  \BibitemOpen
  \bibfield{author}{%
  \bibinfo {author} {\bibfnamefont{P.}~\bibnamefont{Meade}}\ and\ \bibinfo
  {author} {\bibfnamefont{L.}~\bibnamefont{Randall}},\ }%
  \bibfield{journal}{%
  \Doi{10.1088/1126-6708/2008/05/003}{\bibinfo {journal} {JHEP}}\ }%
  \textbf{\bibinfo {volume} {0805}},\ \bibinfo {pages} {003} (\bibinfo {year}
  {2008}),\ \Eprint{http://arxiv.org/abs/0708.3017}{arXiv:0708.3017 [hep-ph]}%
  \bibAnnoteFile{NoStop}{Meade:2007sz}%
\bibitem{Dimopoulos:2001hw}%
  \BibitemOpen
  \bibfield{author}{%
  \bibinfo {author} {\bibfnamefont{S.}~\bibnamefont{Dimopoulos}}\ and\ \bibinfo
  {author} {\bibfnamefont{G.~L.}\ \bibnamefont{Landsberg}},\ }%
  \bibfield{journal}{%
  \Doi{10.1103/PhysRevLett.87.161602}{\bibinfo {journal} {Phys.Rev.Lett.}}\ }%
  \textbf{\bibinfo {volume} {87}},\ \bibinfo {pages} {161602} (\bibinfo {year}
  {2001}),\ \Eprint{http://arxiv.org/abs/hep-ph/0106295}{arXiv:hep-ph/0106295
  [hep-ph]}%
  \bibAnnoteFile{NoStop}{Dimopoulos:2001hw}%
\bibitem{Banks:1999gd}%
  \BibitemOpen
  \bibfield{author}{%
  \bibinfo {author} {\bibfnamefont{T.}~\bibnamefont{Banks}}\ and\ \bibinfo
  {author} {\bibfnamefont{W.}~\bibnamefont{Fischler}}}%
   (\bibinfo {year} {1999}),\
  \Eprint{http://arxiv.org/abs/hep-th/9906038}{arXiv:hep-th/9906038 [hep-th]}%
  \bibAnnoteFile{NoStop}{Banks:1999gd}%
\bibitem{Giddings:2001bu}%
  \BibitemOpen
  \bibfield{author}{%
  \bibinfo {author} {\bibfnamefont{S.~B.}\ \bibnamefont{Giddings}}\ and\
  \bibinfo {author} {\bibfnamefont{S.~D.}\ \bibnamefont{Thomas}},\ }%
  \bibfield{journal}{%
  \Doi{10.1103/PhysRevD.65.056010}{\bibinfo {journal} {Phys.Rev.}}\ }%
  \textbf{\bibinfo {volume} {D65}},\ \bibinfo {pages} {056010} (\bibinfo {year}
  {2002}),\ \Eprint{http://arxiv.org/abs/hep-ph/0106219}{arXiv:hep-ph/0106219
  [hep-ph]}%
  \bibAnnoteFile{NoStop}{Giddings:2001bu}%
\bibitem{Feng:2001ib}%
  \BibitemOpen
  \bibfield{author}{%
  \bibinfo {author} {\bibfnamefont{J.~L.}\ \bibnamefont{Feng}}\ and\ \bibinfo
  {author} {\bibfnamefont{A.~D.}\ \bibnamefont{Shapere}},\ }%
  \bibfield{journal}{%
  \Doi{10.1103/PhysRevLett.88.021303}{\bibinfo {journal} {Phys.Rev.Lett.}}\ }%
  \textbf{\bibinfo {volume} {88}},\ \bibinfo {pages} {021303} (\bibinfo {year}
  {2002}),\ \Eprint{http://arxiv.org/abs/hep-ph/0109106}{arXiv:hep-ph/0109106
  [hep-ph]}%
  \bibAnnoteFile{NoStop}{Feng:2001ib}%
\bibitem{Anchordoqui:2003ug}%
  \BibitemOpen
  \bibfield{author}{%
  \bibinfo {author} {\bibfnamefont{L.~A.}\ \bibnamefont{Anchordoqui}}, \bibinfo
  {author} {\bibfnamefont{J.~L.}\ \bibnamefont{Feng}}, \bibinfo {author}
  {\bibfnamefont{H.}~\bibnamefont{Goldberg}},\ and\ \bibinfo {author}
  {\bibfnamefont{A.~D.}\ \bibnamefont{Shapere}},\ }%
  \bibfield{journal}{%
  \Doi{10.1016/j.physletb.2004.05.051}{\bibinfo {journal} {Phys.Lett.}}\ }%
  \textbf{\bibinfo {volume} {B594}},\ \bibinfo {pages} {363} (\bibinfo {year}
  {2004}),\ \Eprint{http://arxiv.org/abs/hep-ph/0311365}{arXiv:hep-ph/0311365
  [hep-ph]}%
  \bibAnnoteFile{NoStop}{Anchordoqui:2003ug}%
\bibitem{Anchordoqui:2001cg}%
  \BibitemOpen
  \bibfield{author}{%
  \bibinfo {author} {\bibfnamefont{L.~A.}\ \bibnamefont{Anchordoqui}}, \bibinfo
  {author} {\bibfnamefont{J.~L.}\ \bibnamefont{Feng}}, \bibinfo {author}
  {\bibfnamefont{H.}~\bibnamefont{Goldberg}},\ and\ \bibinfo {author}
  {\bibfnamefont{A.~D.}\ \bibnamefont{Shapere}},\ }%
  \bibfield{journal}{%
  \Doi{10.1103/PhysRevD.65.124027}{\bibinfo {journal} {Phys.Rev.}}\ }%
  \textbf{\bibinfo {volume} {D65}},\ \bibinfo {pages} {124027} (\bibinfo {year}
  {2002}),\ \Eprint{http://arxiv.org/abs/hep-ph/0112247}{arXiv:hep-ph/0112247
  [hep-ph]}%
  \bibAnnoteFile{NoStop}{Anchordoqui:2001cg}%
\bibitem{Anchordoqui:2003jr}%
  \BibitemOpen
  \bibfield{author}{%
  \bibinfo {author} {\bibfnamefont{L.~A.}\ \bibnamefont{Anchordoqui}}, \bibinfo
  {author} {\bibfnamefont{J.~L.}\ \bibnamefont{Feng}}, \bibinfo {author}
  {\bibfnamefont{H.}~\bibnamefont{Goldberg}},\ and\ \bibinfo {author}
  {\bibfnamefont{A.~D.}\ \bibnamefont{Shapere}},\ }%
  \bibfield{journal}{%
  \Doi{10.1103/PhysRevD.68.104025}{\bibinfo {journal} {Phys.Rev.}}\ }%
  \textbf{\bibinfo {volume} {D68}},\ \bibinfo {pages} {104025} (\bibinfo {year}
  {2003}),\ \Eprint{http://arxiv.org/abs/hep-ph/0307228}{arXiv:hep-ph/0307228
  [hep-ph]}%
  \bibAnnoteFile{NoStop}{Anchordoqui:2003jr}%
\bibitem{Kowalski:2002gb}%
  \BibitemOpen
  \bibfield{author}{%
  \bibinfo {author} {\bibfnamefont{M.}~\bibnamefont{Kowalski}}, \bibinfo
  {author} {\bibfnamefont{A.}~\bibnamefont{Ringwald}},\ and\ \bibinfo {author}
  {\bibfnamefont{H.}~\bibnamefont{Tu}},\ }%
  \bibfield{journal}{%
  \Doi{10.1016/S0370-2693(02)01235-2}{\bibinfo {journal} {Phys.Lett.}}\ }%
  \textbf{\bibinfo {volume} {B529}},\ \bibinfo {pages} {1} (\bibinfo {year}
  {2002}),\ \Eprint{http://arxiv.org/abs/hep-ph/0201139}{arXiv:hep-ph/0201139
  [hep-ph]}%
  \bibAnnoteFile{NoStop}{Kowalski:2002gb}%
\bibitem{Ringwald:2001vk}%
  \BibitemOpen
  \bibfield{author}{%
  \bibinfo {author} {\bibfnamefont{A.}~\bibnamefont{Ringwald}}\ and\ \bibinfo
  {author} {\bibfnamefont{H.}~\bibnamefont{Tu}},\ }%
  \bibfield{journal}{%
  \Doi{10.1016/S0370-2693(01)01421-6}{\bibinfo {journal} {Phys.Lett.}}\ }%
  \textbf{\bibinfo {volume} {B525}},\ \bibinfo {pages} {135} (\bibinfo {year}
  {2002}),\ \Eprint{http://arxiv.org/abs/hep-ph/0111042}{arXiv:hep-ph/0111042
  [hep-ph]}%
  \bibAnnoteFile{NoStop}{Ringwald:2001vk}%
\bibitem{Calmet:2008dg}%
  \BibitemOpen
  \bibfield{author}{%
  \bibinfo {author} {\bibfnamefont{X.}~\bibnamefont{Calmet}}, \bibinfo {author}
  {\bibfnamefont{W.}~\bibnamefont{Gong}},\ and\ \bibinfo {author}
  {\bibfnamefont{S.~D.}\ \bibnamefont{Hsu}},\ }%
  \bibfield{journal}{%
  \Doi{10.1016/j.physletb.2008.08.011}{\bibinfo {journal} {Phys.Lett.}}\ }%
  \textbf{\bibinfo {volume} {B668}},\ \bibinfo {pages} {20} (\bibinfo {year}
  {2008}),\ \Eprint{http://arxiv.org/abs/0806.4605}{arXiv:0806.4605 [hep-ph]}%
  \bibAnnoteFile{NoStop}{Calmet:2008dg}%
\bibitem{Calmet:2011ta}%
  \BibitemOpen
  \bibfield{author}{%
  \bibinfo {author} {\bibfnamefont{X.}~\bibnamefont{Calmet}}, \bibinfo {author}
  {\bibfnamefont{D.}~\bibnamefont{Fragkakis}},\ and\ \bibinfo {author}
  {\bibfnamefont{N.}~\bibnamefont{Gausmann}},\ }%
  \bibfield{journal}{%
  \Doi{10.1140/epjc/s10052-011-1781-4}{\bibinfo {journal} {Eur.Phys.J.}}\ }%
  \textbf{\bibinfo {volume} {C71}},\ \bibinfo {pages} {1781} (\bibinfo {year}
  {2011}),\ \Eprint{http://arxiv.org/abs/1105.1779}{arXiv:1105.1779 [hep-ph]}%
  \bibAnnoteFile{NoStop}{Calmet:2011ta}%
\bibitem{Calmet:2012cn}%
  \BibitemOpen
  \bibfield{author}{%
  \bibinfo {author} {\bibfnamefont{X.}~\bibnamefont{{Calmet}}}, \bibinfo
  {author} {\bibfnamefont{D.}~\bibnamefont{{Fragkakis}}},\ and\ \bibinfo
  {author} {\bibfnamefont{N.}~\bibnamefont{{Gausmann}}},\ }%
  \bibfield{journal}{%
  \bibinfo {journal} {Chap. 8 in A.J. Bauer and D.G.Eiffel editors,Black Holes:
  Evolution, Theory and Thermodynamics Nova Publishers, New York, 2012}}%
   (\bibinfo {month} {Jan.}\ \bibinfo {year} {2012}),\
  \Eprint{http://arxiv.org/abs/1201.4463}{arXiv:1201.4463 [hep-ph]}%
  \bibAnnoteFile{NoStop}{Calmet:2012cn}%
\bibitem{Cavaglia:2002si}%
  \BibitemOpen
  \bibfield{author}{%
  \bibinfo {author} {\bibfnamefont{M.}~\bibnamefont{Cavaglia}},\ }%
  \bibfield{journal}{%
  \Doi{10.1142/S0217751X03013569}{\bibinfo {journal} {Int.J.Mod.Phys.}}\ }%
  \textbf{\bibinfo {volume} {A18}},\ \bibinfo {pages} {1843} (\bibinfo {year}
  {2003}),\ \Eprint{http://arxiv.org/abs/hep-ph/0210296}{arXiv:hep-ph/0210296
  [hep-ph]}%
  \bibAnnoteFile{NoStop}{Cavaglia:2002si}%
\bibitem{Kanti:2004nr}%
  \BibitemOpen
  \bibfield{author}{%
  \bibinfo {author} {\bibfnamefont{P.}~\bibnamefont{Kanti}},\ }%
  \bibfield{journal}{%
  \Doi{10.1142/S0217751X04018324}{\bibinfo {journal} {Int.J.Mod.Phys.}}\ }%
  \textbf{\bibinfo {volume} {A19}},\ \bibinfo {pages} {4899} (\bibinfo {year}
  {2004}),\ \Eprint{http://arxiv.org/abs/hep-ph/0402168}{arXiv:hep-ph/0402168
  [hep-ph]}%
  \bibAnnoteFile{NoStop}{Kanti:2004nr}%
\bibitem{Calmet:2008rv}%
  \BibitemOpen
  \bibfield{author}{%
  \bibinfo {author} {\bibfnamefont{X.}~\bibnamefont{Calmet}}\ and\ \bibinfo
  {author} {\bibfnamefont{M.}~\bibnamefont{Feliciangeli}},\ }%
  \bibfield{journal}{%
  \Doi{10.1103/PhysRevD.78.067702}{\bibinfo {journal} {Phys.Rev.}}\ }%
  \textbf{\bibinfo {volume} {D78}},\ \bibinfo {pages} {067702} (\bibinfo {year}
  {2008}),\ \Eprint{http://arxiv.org/abs/0806.4304}{arXiv:0806.4304 [hep-ph]}%
  \bibAnnoteFile{NoStop}{Calmet:2008rv}%
\bibitem{Calmet:2012mf}%
  \BibitemOpen
  \bibfield{author}{%
  \bibinfo {author} {\bibfnamefont{X.}~\bibnamefont{Calmet}}, \bibinfo {author}
  {\bibfnamefont{L.~I.}\ \bibnamefont{Caramete}},\ and\ \bibinfo {author}
  {\bibfnamefont{O.}~\bibnamefont{Micu}},\ }%
  \bibfield{journal}{%
  \Doi{10.1007/JHEP11(2012)104}{\bibinfo {journal} {JHEP}}\ }%
  \textbf{\bibinfo {volume} {1211}},\ \bibinfo {pages} {104} (\bibinfo {year}
  {2012}),\ \Eprint{http://arxiv.org/abs/1204.2520}{arXiv:1204.2520 [hep-ph]}%
  \bibAnnoteFile{NoStop}{Calmet:2012mf}%
\bibitem{2011NIMPA.656...11A}%
  \BibitemOpen
  \bibfield{author}{%
  \bibinfo {author} {\bibfnamefont{M.}~\bibnamefont{{Ageron}}}, \bibinfo
  {author} {\bibfnamefont{J.~A.}\ \bibnamefont{{Aguilar}}}, \bibinfo {author}
  {\bibfnamefont{I.}~\bibnamefont{{Al Samarai}}}, \bibinfo {author}
  {\bibfnamefont{A.}~\bibnamefont{{Albert}}}, \bibinfo {author}
  {\bibfnamefont{F.}~\bibnamefont{{Ameli}}}, \bibinfo {author}
  {\bibfnamefont{M.}~\bibnamefont{{Andr{\'e}}}}, \bibinfo {author}
  {\bibfnamefont{M.}~\bibnamefont{{Anghinolfi}}}, \bibinfo {author}
  {\bibfnamefont{G.}~\bibnamefont{{Anton}}}, \bibinfo {author}
  {\bibfnamefont{S.}~\bibnamefont{{Anvar}}}, \bibinfo {author}
  {\bibfnamefont{M.}~\bibnamefont{{Ardid}}},\ and\ \bibinfo {author}
  {\bibnamefont{et~al.}},\ }%
  \bibfield{journal}{%
  \Doi{10.1016/j.nima.2011.06.103}{\bibinfo {journal} {Nuclear Instruments and
  Methods in Physics Research A}}\ }%
  \textbf{\bibinfo {volume} {656}},\ \bibinfo {pages} {11} (\bibinfo {month}
  {Nov.}\ \bibinfo {year} {2011}),\
  \Eprint{http://arxiv.org/abs/1104.1607}{arXiv:1104.1607 [astro-ph.IM]}%
  \bibAnnoteFile{NoStop}{2011NIMPA.656...11A}%
\bibitem{2006APh....26..155I}%
  \BibitemOpen
  \bibfield{author}{%
  \bibinfo {author} {\bibnamefont{{IceCube Collaboration}}}, \bibinfo {author}
  {\bibfnamefont{A.}~\bibnamefont{{Achterberg}}}, \bibinfo {author}
  {\bibfnamefont{M.}~\bibnamefont{{Ackermann}}}, \bibinfo {author}
  {\bibfnamefont{J.}~\bibnamefont{{Adams}}}, \bibinfo {author}
  {\bibfnamefont{J.}~\bibnamefont{{Ahrens}}}, \bibinfo {author}
  {\bibfnamefont{K.}~\bibnamefont{{Andeen}}}, \bibinfo {author}
  {\bibfnamefont{D.~W.}\ \bibnamefont{{Atlee}}}, \bibinfo {author}
  {\bibfnamefont{J.}~\bibnamefont{{Baccus}}}, \bibinfo {author}
  {\bibfnamefont{J.~N.}\ \bibnamefont{{Bahcall}}}, \bibinfo {author}
  {\bibfnamefont{X.}~\bibnamefont{{Bai}}},\ and\ \bibinfo {author}
  {\bibnamefont{et~al.}},\ }%
  \bibfield{journal}{%
  \Doi{10.1016/j.astropartphys.2006.06.007}{\bibinfo {journal} {Astroparticle
  Physics}}\ }%
  \textbf{\bibinfo {volume} {26}},\ \bibinfo {pages} {155} (\bibinfo {month}
  {Oct.}\ \bibinfo {year} {2006}),\
  \Eprint{http://arxiv.org/abs/arXiv:astro-ph/0604450}{arXiv:astro-ph/0604450}%
  \bibAnnoteFile{NoStop}{2006APh....26..155I}%
\bibitem{Gandhi:1995tf}%
  \BibitemOpen
  \bibfield{author}{%
  \bibinfo {author} {\bibfnamefont{R.}~\bibnamefont{Gandhi}}, \bibinfo {author}
  {\bibfnamefont{C.}~\bibnamefont{Quigg}}, \bibinfo {author}
  {\bibfnamefont{M.~H.}\ \bibnamefont{Reno}},\ and\ \bibinfo {author}
  {\bibfnamefont{I.}~\bibnamefont{Sarcevic}},\ }%
  \bibfield{journal}{%
  \Doi{10.1016/0927-6505(96)00008-4}{\bibinfo {journal} {Astropart.Phys.}}\ }%
  \textbf{\bibinfo {volume} {5}},\ \bibinfo {pages} {81} (\bibinfo {year}
  {1996}),\ \Eprint{http://arxiv.org/abs/hep-ph/9512364}{arXiv:hep-ph/9512364
  [hep-ph]}%
  \bibAnnoteFile{NoStop}{Gandhi:1995tf}%
\bibitem{Feng:2001ue}%
  \BibitemOpen
  \bibfield{author}{%
  \bibinfo {author} {\bibfnamefont{J.~L.}\ \bibnamefont{Feng}}, \bibinfo
  {author} {\bibfnamefont{P.}~\bibnamefont{Fisher}}, \bibinfo {author}
  {\bibfnamefont{F.}~\bibnamefont{Wilczek}},\ and\ \bibinfo {author}
  {\bibfnamefont{T.~M.}\ \bibnamefont{Yu}},\ }%
  \bibfield{journal}{%
  \Doi{10.1103/PhysRevLett.88.161102}{\bibinfo {journal} {Phys.Rev.Lett.}}\ }%
  \textbf{\bibinfo {volume} {88}},\ \bibinfo {pages} {161102} (\bibinfo {year}
  {2002}),\ \Eprint{http://arxiv.org/abs/hep-ph/0105067}{arXiv:hep-ph/0105067
  [hep-ph]}%
  \bibAnnoteFile{NoStop}{Feng:2001ue}%
\bibitem{Yoshino:2002br}%
  \BibitemOpen
  \bibfield{author}{%
  \bibinfo {author} {\bibfnamefont{H.}~\bibnamefont{Yoshino}}\ and\ \bibinfo
  {author} {\bibfnamefont{Y.}~\bibnamefont{Nambu}},\ }%
  \bibfield{journal}{%
  \Doi{10.1103/PhysRevD.66.065004}{\bibinfo {journal} {Phys.Rev.}}\ }%
  \textbf{\bibinfo {volume} {D66}},\ \bibinfo {pages} {065004} (\bibinfo {year}
  {2002}),\ \Eprint{http://arxiv.org/abs/gr-qc/0204060}{arXiv:gr-qc/0204060
  [gr-qc]}%
  \bibAnnoteFile{NoStop}{Yoshino:2002br}%
\bibitem{Stojkovic:2005fx}%
  \BibitemOpen
  \bibfield{author}{%
  \bibinfo {author} {\bibfnamefont{D.}~\bibnamefont{Stojkovic}}, \bibinfo
  {author} {\bibfnamefont{G.~D.}\ \bibnamefont{Starkman}},\ and\ \bibinfo
  {author} {\bibfnamefont{D.-C.}\ \bibnamefont{Dai}},\ }%
  \bibfield{journal}{%
  \Doi{10.1103/PhysRevLett.96.041303}{\bibinfo {journal} {Phys.Rev.Lett.}}\ }%
  \textbf{\bibinfo {volume} {96}},\ \bibinfo {pages} {041303} (\bibinfo {year}
  {2006}),\ \Eprint{http://arxiv.org/abs/hep-ph/0505112}{arXiv:hep-ph/0505112
  [hep-ph]}%
  \bibAnnoteFile{NoStop}{Stojkovic:2005fx}%
\bibitem{Olinto:2000sa}%
  \BibitemOpen
  \bibfield{author}{%
  \bibinfo {author} {\bibfnamefont{A.}~\bibnamefont{Olinto}},\ }%
  \bibfield{journal}{%
  \Doi{10.1016/S0370-1573(00)00028-4}{\bibinfo {journal} {Phys.Rept.}}\ }%
  \textbf{\bibinfo {volume} {333}},\ \bibinfo {pages} {329} (\bibinfo {year}
  {2000}),\
  \Eprint{http://arxiv.org/abs/astro-ph/0002006}{arXiv:astro-ph/0002006
  [astro-ph]}%
  \bibAnnoteFile{NoStop}{Olinto:2000sa}%
\bibitem{Allard:2006mv}%
  \BibitemOpen
  \bibfield{author}{%
  \bibinfo {author} {\bibfnamefont{D.}~\bibnamefont{Allard}}, \bibinfo {author}
  {\bibfnamefont{M.}~\bibnamefont{Ave}}, \bibinfo {author}
  {\bibfnamefont{N.}~\bibnamefont{Busca}}, \bibinfo {author}
  {\bibfnamefont{M.}~\bibnamefont{Malkan}}, \bibinfo {author}
  {\bibfnamefont{A.}~\bibnamefont{Olinto}}, \emph{et~al.},\ }%
  \bibfield{journal}{%
  \Doi{10.1088/1475-7516/2006/09/005}{\bibinfo {journal} {JCAP}}\ }%
  \textbf{\bibinfo {volume} {0609}},\ \bibinfo {pages} {005} (\bibinfo {year}
  {2006}),\
  \Eprint{http://arxiv.org/abs/astro-ph/0605327}{arXiv:astro-ph/0605327
  [astro-ph]}%
  \bibAnnoteFile{NoStop}{Allard:2006mv}%
\bibitem{Hooper:2004jc}%
  \BibitemOpen
  \bibfield{author}{%
  \bibinfo {author} {\bibfnamefont{D.}~\bibnamefont{Hooper}}, \bibinfo {author}
  {\bibfnamefont{A.}~\bibnamefont{Taylor}},\ and\ \bibinfo {author}
  {\bibfnamefont{S.}~\bibnamefont{Sarkar}},\ }%
  \bibfield{journal}{%
  \Doi{10.1016/j.astropartphys.2004.11.002}{\bibinfo {journal}
  {Astropart.Phys.}}\ }%
  \textbf{\bibinfo {volume} {23}},\ \bibinfo {pages} {11} (\bibinfo {year}
  {2005}),\
  \Eprint{http://arxiv.org/abs/astro-ph/0407618}{arXiv:astro-ph/0407618
  [astro-ph]}%
  \bibAnnoteFile{NoStop}{Hooper:2004jc}%
\bibitem{2013arXiv13011703K}%
  \BibitemOpen
  \bibfield{author}{%
  \bibinfo {author} {\bibfnamefont{M.~D.}\ \bibnamefont{{Kistler}}}, \bibinfo
  {author} {\bibfnamefont{T.}~\bibnamefont{{Stanev}}},\ and\ \bibinfo {author}
  {\bibfnamefont{H.}~\bibnamefont{{Yuksel}}},\ }%
  \bibfield{journal}{%
  \bibinfo {journal} {ArXiv e-prints}}%
   (\bibinfo {month} {Jan.}\ \bibinfo {year} {2013}),\
  \Eprint{http://arxiv.org/abs/1301.1703}{arXiv:1301.1703 [astro-ph.HE]}%
  \bibAnnoteFile{NoStop}{2013arXiv13011703K}%
\bibitem{2001PhRvD63b3003M}%
  \BibitemOpen
  \bibfield{author}{%
  \bibinfo {author} {\bibfnamefont{K.}~\bibnamefont{{Mannheim}}}, \bibinfo
  {author} {\bibfnamefont{R.~J.}\ \bibnamefont{{Protheroe}}},\ and\ \bibinfo
  {author} {\bibfnamefont{J.~P.}\ \bibnamefont{{Rachen}}},\ }%
  \bibfield{journal}{%
  \Doi{10.1103/PhysRevD.63.023003}{\bibinfo {journal} {\prd}}\ }%
  \textbf{\bibinfo {volume} {63}},\ \bibinfo {eid} {023003} (\bibinfo {month}
  {Jan.}\ \bibinfo {year} {2001}),\
  \Eprint{http://arxiv.org/abs/arXiv:astro-ph/9812398}{arXiv:astro-ph/9812398}%
  \bibAnnoteFile{NoStop}{2001PhRvD63b3003M}%
\bibitem{2013NIMPA700188I}%
  \BibitemOpen
  \bibfield{author}{%
  \bibinfo {author} {\bibnamefont{{IceCube Collaboration}}}, \bibinfo {author}
  {\bibfnamefont{R.}~\bibnamefont{{Abbasi}}}, \bibinfo {author}
  {\bibfnamefont{Y.}~\bibnamefont{{Abdou}}}, \bibinfo {author}
  {\bibfnamefont{M.}~\bibnamefont{{Ackermann}}}, \bibinfo {author}
  {\bibfnamefont{J.}~\bibnamefont{{Adams}}}, \bibinfo {author}
  {\bibfnamefont{J.~A.}\ \bibnamefont{{Aguilar}}}, \bibinfo {author}
  {\bibfnamefont{M.}~\bibnamefont{{Ahlers}}}, \bibinfo {author}
  {\bibfnamefont{D.}~\bibnamefont{{Altmann}}}, \bibinfo {author}
  {\bibfnamefont{K.}~\bibnamefont{{Andeen}}}, \bibinfo {author}
  {\bibfnamefont{J.}~\bibnamefont{{Auffenberg}}},\ and\ \bibinfo {author}
  {\bibnamefont{et~al.}},\ }%
  \bibfield{journal}{%
  \Doi{10.1016/j.nima.2012.10.067}{\bibinfo {journal} {Nuclear Instruments and
  Methods in Physics Research A}}\ }%
  \textbf{\bibinfo {volume} {700}},\ \bibinfo {pages} {188} (\bibinfo {month}
  {Feb.}\ \bibinfo {year} {2013}),\
  \Eprint{http://arxiv.org/abs/1207.6326}{arXiv:1207.6326 [astro-ph.IM]}%
  \bibAnnoteFile{NoStop}{2013NIMPA700188I}%
\bibitem{Calmet:2012fv}%
  \BibitemOpen
  \bibfield{author}{%
  \bibinfo {author} {\bibfnamefont{X.}~\bibnamefont{Calmet}}\ and\ \bibinfo
  {author} {\bibfnamefont{N.}~\bibnamefont{Gausmann}}}%
   (\bibinfo {year} {2012}),\
  \Eprint{http://arxiv.org/abs/1209.4618}{arXiv:1209.4618 [hep-ph]}%
  \bibAnnoteFile{NoStop}{Calmet:2012fv}%
\bibitem{Note3}%
  \BibitemOpen
  \bibinfo {note} {This version of CORSIKA was obtained changing the medium in
  which the simulation takes place from air to ice. It is maintained by J.
  Bolmont in DESY at Zeuthen. This work was inspired by the work done by T.
  Sloan (Lancaster University) for the ACoRNE collaboration (Astropart. Phys.,
  28, 366 (2007)).}%
  \bibAnnoteFile{Stop}{Note3}%
\bibitem{corsika}%
  \BibitemOpen
  \bibfield{author}{%
  \bibinfo {author} {\bibfnamefont{D.}~\bibnamefont{Heck}}\ and\ \bibinfo
  {author} {\bibfnamefont{J.}~\bibnamefont{Knapp}},\ }%
  \bibfield{journal}{%
  \bibinfo {journal} {{Report {\bf FZKA 6097} (1998), Forschungszentrum
  Karlsruhe; available from http://www-ik.fzk.de/\textasciitilde
  heck/publications/}}}%
   (\bibinfo {year} {1989})%
  \bibAnnoteFile{NoStop}{corsika}%
\bibitem{corsika1}%
  \BibitemOpen
  \bibfield{author}{%
  \bibinfo {author} {\bibfnamefont{D.}~\bibnamefont{Heck}}, \bibinfo {author}
  {\bibfnamefont{J.}~\bibnamefont{Knapp}}, \bibinfo {author}
  {\bibfnamefont{J.}~\bibnamefont{Capdevielle}}, \bibinfo {author}
  {\bibfnamefont{G.}~\bibnamefont{Schatz}},\ and\ \bibinfo {author}
  {\bibfnamefont{T.}~\bibnamefont{Thouw}},\ }%
  \bibfield{journal}{%
  \bibinfo {journal} {{Report {\bf FZKA 6019} (1998), Forschungszentrum
  Karlsruhe; available from http://www-ik.fzk.de/corsika/physics$\_
  $description/corsika$\_ $phys.html}}}%
   (\bibinfo {year} {1998})%
  \bibAnnoteFile{NoStop}{corsika1}%
\bibitem{1997NuPhS..52...17K}%
  \BibitemOpen
  \bibfield{author}{%
  \bibinfo {author} {\bibfnamefont{N.~N.}\ \bibnamefont{{Kalmykov}}}, \bibinfo
  {author} {\bibfnamefont{S.~S.}\ \bibnamefont{{Ostapchenko}}},\ and\ \bibinfo
  {author} {\bibfnamefont{A.~I.}\ \bibnamefont{{Pavlov}}},\ }%
  \bibfield{journal}{%
  \Doi{10.1016/S0920-5632(96)00846-8}{\bibinfo {journal} {Nuclear Physics B
  Proceedings Supplements}}\ }%
  \textbf{\bibinfo {volume} {52}},\ \bibinfo {pages} {17} (\bibinfo {month}
  {Feb.}\ \bibinfo {year} {1997})%
  \bibAnnoteFile{NoStop}{1997NuPhS..52...17K}%
\bibitem{Karle:2010xx}%
  \BibitemOpen
  \bibfield{author}{%
  \bibinfo {author} {\bibfnamefont{A.}~\bibnamefont{Karle}} (\bibinfo
  {collaboration} {IceCube Collaboration})}%
   (\bibinfo {year} {2010}),\
  \Eprint{http://arxiv.org/abs/1003.5715}{arXiv:1003.5715 [astro-ph.HE]}%
  \bibAnnoteFile{NoStop}{Karle:2010xx}%
\bibitem{KM3NeT}%
  \BibitemOpen
  \
  \Eprint{http://arxiv.org/abs/{http://www.km3net.org/TDR/TDRKM3NeT.pdf}}{{htt%
p://www.km3net.org/TDR/TDRKM3NeT.pdf}}%
  \bibAnnoteFile{NoStop}{KM3NeT}%
\bibitem{Aartsen:2013dla}%
  \BibitemOpen
  \bibfield{author}{%
  \bibinfo {author} {\bibfnamefont{M.}~\bibnamefont{Aartsen}} \emph{et~al.}
  (\bibinfo {collaboration} {The IceCube Collaboration})}%
   (\bibinfo {year} {2013}),\
  \Eprint{http://arxiv.org/abs/1309.6979}{arXiv:1309.6979 [astro-ph.HE]}%
  \bibAnnoteFile{NoStop}{Aartsen:2013dla}%
\bibitem{2013NIMPA.725...13K}%
  \BibitemOpen
  \bibfield{author}{%
  \bibinfo {author} {\bibfnamefont{P.}~\bibnamefont{{Kooijman}}}\ and\ \bibinfo
  {author} {\bibnamefont{{KM3NeT Consortium}}},\ }%
  \bibfield{journal}{%
  \Doi{10.1016/j.nima.2012.12.055}{\bibinfo {journal} {Nuclear Instruments and
  Methods in Physics Research A}}\ }%
  \textbf{\bibinfo {volume} {725}},\ \bibinfo {pages} {13} (\bibinfo {month}
  {Oct.}\ \bibinfo {year} {2013})%
  \bibAnnoteFile{NoStop}{2013NIMPA.725...13K}%
\end{thebibliography}%

\end{document}